\documentclass[12pt]{article}
\usepackage{amsfonts, amsmath, amsthm, amssymb, graphicx, latexsym, slashbox, natbib, bm, color, threeparttable}

\usepackage[left=1in,top=1in,right=1in,bottom=1in,nohead,paperwidth=8.5in, paperheight=11in]{geometry} 
\title{\textbf{Sparse cointegration}}
\author{Ines Wilms\thanks{Corresponding author: Faculty of Economics and Business, KU Leuven, Naamsestraat 69, B-3000 Leuven, Belgium. E-mail address: ines.wilms@kuleuven.be.}, Christophe Croux \\
\it Faculty of Economics and Business, KU Leuven, Belgium}
\date{}

\begin{document}
\maketitle

\noindent {\it Abstract.} Cointegration analysis is used to estimate the long-run equilibrium relations between several time series. The coefficients of these long-run equilibrium relations are the cointegrating vectors. In this paper, we provide a sparse estimator of the cointegrating vectors. The estimation technique is sparse in the sense that some elements of the cointegrating vectors will be estimated as zero. For this purpose, we combine a penalized estimation procedure for vector autoregressive models with  sparse reduced rank regression. The sparse cointegration procedure achieves a higher estimation accuracy than the traditional Johansen cointegration approach in settings where the true cointegrating vectors have a sparse structure, and/or when the sample size is low compared to the number of time series. We also discuss a criterion to determine the cointegration rank and we illustrate its good performance in several simulation settings. In a first empirical application we investigate whether the expectations hypothesis of the term structure of interest rates, implying sparse cointegrating vectors, holds in practice. In a second empirical application we show that forecast performance in high-dimensional systems can be improved by sparsely estimating the cointegration relations.

\bigskip

\noindent {\it Keyswords.} Adaptive lasso; Penalized estimation; Reduced rank regression; Sparse estimation;  Vector error correcting model

\newpage

\section{Introduction}
High-dimensional datasets containing thousands of economic time series are commonly available and accessible at reasonable cost \citep{Stock02, Clements08, Fan11}. The aim of this paper is to develop a cointegration technique for high-dimensional time series. In a cointegration analysis, long-run equilibrium relations, often implied by economic theory, are estimated. In financial economics, for instance, cointegration analysis is used to investigate whether the expectations hypothesis of the term structure of interest rates (EHT) holds in practice. The Vector Error Correcting Model (VECM) (see e.g. \citealp{Lutkepohl07}) is used to estimate and test for the cointegration relationships. Various approaches to test for cointegration are existing (see  among others \citealp{Engle87,Phillips90}), among which the system cointegration test of \cite{Johansen88} has become most popular. 

The conventional Johansen system cointegration approach has, however, some limitations. In high-dimensional settings, where the number of time series is large compared to the sample size, the estimation imprecision will be large. Johansen's approach is based on the estimation of a Vector AutoRegressive (VAR) model and a canonical correlation analysis. A drawback of the VAR model is that the number of parameters increases quadratically with the number of included time series. Consequently, regression parameters are estimated inaccurately if only a limited number of observations is available.  When the number of time series exceeds the sample size, Johansen's approach can not even be applied. 

In this paper, we introduce a penalized maximum likelihood approach  to estimate the cointegrating vectors in a sparse way, i.e. some of its components are estimated as exactly zero.  
Sparse estimation techniques show good performance in various fields, such as, for instance, economics (e.g. \citealp{Fan11}),  econometrics (e.g. \citealp{Caner14}), or macro-economics (e.g. \citealp{Korobilis13}). 
A sparse cointegration approach is useful for several reasons. First, a sparse approach is justified if economic theory implies sparsity in the cointegrating vectors (as is the case for the EHT, see e.g. \citealp{Engsted94}). 
Secondly, a sparse approach facilitates model interpretation since only a limited number of variables, those corresponding to the non-zero coefficients, enter the estimated long-run equilibrium relations. Thirdly, sparsity improves forecast performance through variance reduction. Lastly, the sparse cointegration technique, in contrast to Johansen's method, can be applied when the number of time series exceeds the sample size. We show in a simulation study that the sparse cointegration technique significantly outperforms Johansen's approach when the cointegrating vectors have a sparse structure or when the number of time series is large compared to the sample size.

We apply the sparse cointegration technique on a financial and macro-economic dataset. In the first empirical application, we investigate whether the expectations hypothesis of the term structure of interest rates (EHT) holds in practice. Previous research on the validity of the EHT reports evidence in support of the theory at the short end of the term structure (e.g. \citealp{Hall92}; \citealp{Lasak14}). The theory is generally rejected at the longer end (e.g. \citealp{Shea92}; \citealp{Zhang93}; \citealp{Carstensen03}). We test the cointegration implications linked to the EHT for five US interest rates. Using the sparse cointegration technique, we find evidence in favor of the zero-sum restriction (i.e. for each cointegrating vector, the sum of the cointegration coefficients should be equal to zero). In a second empirical application, we use the VECM to perform a forecast exercise in a high-dimensional setting containing a large number of industrial production time series. We  show that sparsely estimating the cointegrating vectors leads to an improvement in forecast performance.

Cointegration analysis in high-dimensions has received little attention in previous research. Large Vector Autoregressive Models, containing a high number of time series relative to the sample size, have been considered extensively. Common approaches are, among others, Dynamic Factor Models (e.g. \citealp{Stock02}), Bayesian VAR Models (e.g. \citealp{Banbura10}) or Reduced-Rank VAR Models (e.g. \citealp{Carriero11, Bernardini14}). Typically, these authors do not account for cointegration. Instead, the time series are either transformed in order to achieve stationarity  (e.g. \citealp{Bernardini14}) or  the (non)-stationarity of the time series is accounted for in the prior distribution of the autoregressive parameters (e.g. \citealp{Banbura10}). Few studies, e.g. \cite{Strachan03} or \cite{Koop06}, do account for cointegration by using a Bayesian method for obtaining estimates of the cointegrating vectors. These Bayesian approaches, in contrast to the sparse cointegration approach discussed in this paper, do not perform variable selection and require prior specification.

The remainder of this article is structured as follows. We describe the sparse cointegration method in Section 2. Section 3 provides more details on the algorithm. Section 4 discusses the Rank Selection Criterion \citep{Bunea2011} to determine the cointegration rank. Section 5 presents the results of a simulation study, Section 6 discusses the findings on the empirical applications. Finally, Section 7 concludes.

\section{Penalized Maximum Likelihood}
Let ${\bf y}_t$ be a $q$-dimensional multivariate time series, where ${\bf y}_t$ is $I(1)$. We assume that ${\bf y}_t$ follows a VAR($p$) model. Any $p^{th}$ order VAR model can be re-written in vector error correcting (VECM) representation \citep{Hamilton91} as follows
\begin{equation} \label{vecm}
{ \bf \Delta  y}_t = \sum_{i=1}^{p-1} { \bf \Gamma}_{i} {\bf \Delta  y}_{t-i}  +  {\bf \Pi} {\bf y}_{t-1}  +  \boldsymbol\varepsilon_t,   \text{\ \ \ \ \ \  }t=1,\ldots,T
\end{equation}
where ${ \bf \Gamma}_{1},\ldots,{ \bf \Gamma}_{p-1}$ are $q \times q$ matrices containing short-run effects, ${ \bf \Pi}$ is a $q \times q$ matrix of rank $r, \  0 \leq r \leq q$ and $\boldsymbol\varepsilon_t$ is assumed to follow a $N_q(\bf{0},\bf{\Sigma})$. 

If we can express ${\bf \Pi} =  \boldsymbol\alpha \boldsymbol\beta^{\prime}$ with $\boldsymbol\alpha$ and $\boldsymbol\beta$ $q \times r$ matrices of full column rank $r$, with $0 < r < q$, then the linear combinations given by $ \boldsymbol\beta^{\prime} {\bf y}_{t} $ are stationary and ${\bf y}_{t}$ is said to be cointegrated with cointegration rank $r$. The cointegrating vectors are the columns of $\boldsymbol\beta$ and the adjustment coefficients the elements of $\boldsymbol\alpha$.

We estimate the model parameters in a penalized maximum likelihood framework. It is convenient to rewrite model \eqref{vecm} in matrix notation:
\begin{equation} \label{vecm_matrix}
{ \bf Y} = { \bf X} { \bf \Gamma} +  { \bf Z} { \bf \Pi}^{\prime} + { \bf E} 
\end{equation}
where ${ \bf Y}=({ \bf \Delta  y}_{p+1},\ldots,{ \bf \Delta  y}_T)^{\prime}$; ${ \bf X}=({ \bf \Delta  X}_{p+1},\ldots,{ \bf \Delta  X}_T)^{\prime}$ with ${ \bf X}_t=({\bf \Delta  y}_{t-1}^{\prime},\ldots,{\bf \Delta  y}_{t-p+1}^{\prime})^{\prime}$; ${ \bf Z}=({\bf y}_{p},\ldots,{\bf y}_{T-1})^{\prime}$; ${ \boldsymbol\Gamma}=({\boldsymbol\Gamma}_1,\ldots,{ \boldsymbol\Gamma}_{p-1})^{\prime}$; and ${ \bf E}=(\boldsymbol\varepsilon_{p+1},\ldots,\boldsymbol\varepsilon_T)^{\prime}$. 
Consider the penalized negative log-likelihood, given by 
\begin{equation}
\mathcal{L}(\boldsymbol\Gamma, \boldsymbol\Pi, \boldsymbol\Omega) = \frac{1}{T}\text{tr} \Bigl( ({ \bf Y} - { \bf X}{ \boldsymbol\Gamma} - { \bf Z}{ \boldsymbol \Pi}^{\prime})\boldsymbol\Omega({ \bf Y} - { \bf X}{ \boldsymbol\Gamma} - { \bf Z}{ \boldsymbol \Pi}^{\prime})^{\prime}\Bigr)  - \text{log}|\boldsymbol\Omega| + \lambda_1 P_1(\boldsymbol\beta) + \lambda_2 P_{2}(\boldsymbol\Gamma) + \lambda_3 P_{3}(\boldsymbol\Omega),  \label{penloglik}
\end{equation}
with $\text{tr}(\cdot)$ denoting the trace,  $\boldsymbol{\Omega}=\boldsymbol\Sigma^{-1}$, and $P_1$, $P_2$ and $P_3$ three penalty functions. 

We use $L_1$ or Lasso penalization \citep{Tibshirani96} on the cointegrating vectors 
\begin{equation}
P_{1}(\boldsymbol\beta)=  \sum_{i=1}^{q}\sum_{j=1}^{r}|\beta_{ij}|. \label{L1penalty}
\end{equation}
As an extension, we also consider the Adaptive Lasso \citep{Zou06} 
\begin{equation}
P_{1}(\boldsymbol\beta)=\sum_{i=1}^{q}\sum_{j=1}^{r}w_{ij}|\beta_{ij}|, \label{L1ALasso}
\end{equation} 
with weights $w_{ij}$. The weights $\hat{w}_{ij}$ are computed as the inverse of the Lasso solution  $\hat{w}_{ij}=1/\hat{\beta}_{ij}^{lasso}$, for $\hat{\beta}_{ij}^{lasso} \neq 0$. The Adaptive Lasso enjoys the oracle property (consistent for variable selection), whereas the Lasso does not \citep{Zou06}.

For the short-run effects $\boldsymbol\Gamma$, we use $L_2$ or Ridge penalization \citep{Hoerl70} 
\begin{equation}
P_{2}(\boldsymbol\Gamma)= \sum_{i=1}^{q}\sum_{j=1}^{q}\sum_{k=1}^{p-1} \gamma_{ijk}^2, \label{L2penalty}
\end{equation}
 with $\gamma_{ijk}$ the $(i,j)th$ element of $\boldsymbol\Gamma_k$. The $L_1$ penalty shrinks parameter estimates towards zero and sets some to exactly zero. Contrary to the $L_1$ penalty, the $L_2$ penalty only shrinks parameter estimates towards zero. We use an $L_2$ penalty for the short-run effects since this is computationally less expensive and we only require sparsity in the cointegrating vectors. Note that using the ridge penalty, estimation remains feasible if the number of time series exceeds the sample size. 
 
Finally, we use $L_1$ penalization for the off-diagonal elements of the  inverse of the error covariance matrix, $\boldsymbol\Omega$, 
\begin{equation}
P_{3}(\boldsymbol\Omega)=  \sum_{k\neq k{\prime}}|\Omega_{kk^{\prime}}|. \label{L1penaltyOmega}
\end{equation} 

The aim is to select $\boldsymbol\Gamma, \boldsymbol\Pi, \boldsymbol\Omega$ so as to minimize \eqref{penloglik} subject to the constraint
\begin{equation}
\boldsymbol\Pi=\boldsymbol\alpha \boldsymbol\beta^{\prime}, \nonumber
\end{equation}
with $\boldsymbol\alpha$ and $\boldsymbol\beta$ $q \times r$ matrices of full column rank $r$. The matrices $\boldsymbol\alpha$ and $\boldsymbol\beta$ are not uniquely defined. Section 3 provides more details on the normalization conditions we impose. For the unpenalized case (i.e. $\lambda_1=0$, $\lambda_2=0$ and $\lambda_3=0$), the objective function \eqref{penloglik} boils down to the one introduced by \cite{Johansen88}. The unpenalized case can be solved either by using an iterative algorithm or by using the closed-form expressions documented in \cite{Johansen88}.

\section{Algorithm}
To find the minimum of the penalized negative log-likelihood in \eqref{penloglik}, we iteratively solve for $\boldsymbol\Gamma$ conditional on $\boldsymbol\Pi,\boldsymbol\Omega$;  for $\boldsymbol\Pi$ conditional on $\boldsymbol\Gamma, \boldsymbol\Omega$; and for $\boldsymbol\Omega$ conditional on $\boldsymbol\Gamma, \boldsymbol\Pi$. 

\bigskip
\noindent
{\bf Solving for $\boldsymbol\Gamma$ conditional on $\boldsymbol\Pi,\boldsymbol\Omega$.}
When $\boldsymbol\Pi$ and $\boldsymbol\Omega$  are fixed, the minimization problem in \eqref{penloglik} is equivalent to minimizing
\begin{equation}
\widehat{\boldsymbol\Gamma}| \boldsymbol\Pi, \boldsymbol\Omega = \underset{\boldsymbol\Gamma}{\operatorname{argmin}} \text{\ \ \ } \frac{1}{T} \text{tr} \Bigl(  ({ \bf Y} - { \bf Z}{ \boldsymbol \Pi}^{\prime} - { \bf X}{ \boldsymbol\Gamma} )\boldsymbol\Omega({ \bf Y} - { \bf Z}{ \boldsymbol \Pi}^{\prime} - { \bf X}{ \boldsymbol\Gamma} )^{\prime}\Bigr) +  \lambda_2 P_{2}(\boldsymbol\Gamma). \label{MLM}
\end{equation}

The above minimization problem is a penalized multivariate regression (see e.g. \citealp{Rothman10}) of $ \left( { \bf Y} - { \bf Z}{ \boldsymbol \Pi}^{\prime} \right )$ on $ { \bf X}$. We solve this penalized multivariate regression using the ridge penalty, as given in equation \eqref{L2penalty}.
The closed-form expression for the estimated short-run dynamics $\widehat{\boldsymbol\Gamma}$ is given by
\begin{equation}
\widehat{\boldsymbol\Gamma} = \left ( {\bf X}^{\text{ridge}\prime} {\bf X}^{\text{ridge}}+ \lambda_2 {\bf I} \right )^{-1} {\bf X}^{\text{ridge}\prime} y^{\text{ridge}}, \nonumber
\end{equation}
with 
\begin{equation}
y^{\text{ridge}}= (\boldsymbol\Omega^{1/2} \otimes \bf I_n )\text{\ vec}\left( { \bf Y} - { \bf Z}{ \boldsymbol \Pi}^{\prime} \right ), \nonumber
\end{equation}
where the latter is a vector of length $nq$ containing the stacked values of the time series given in the columns of the matrix $\left( { \bf Y} - { \bf Z}{ \boldsymbol \Pi}^{\prime} \right )$,   and 
\begin{equation}
{\bf X}^{\text{ridge}} = (\boldsymbol\Omega^{1/2} \otimes \bf I_n ) (\bf I_q \otimes \bf Z), \nonumber
\end{equation}
where $\otimes$ stands for the kronecker product.

\bigskip

\noindent
{\bf Solving for $\boldsymbol\Pi$ conditional on $\boldsymbol\Gamma,\boldsymbol\Omega$.}
When $\boldsymbol\Gamma$ and $\boldsymbol\Omega$ are fixed, the minimization problem in \eqref{penloglik} is equivalent to 
\begin{equation}
(\hat{\boldsymbol\alpha},\hat{\boldsymbol\beta})|\boldsymbol\Gamma_,\boldsymbol\Omega = \underset{\boldsymbol\alpha,\boldsymbol\beta}{\operatorname{argmin}} 
\text{\ \ \ } \frac{1}{T} \text{tr} \Bigl( ({ \bf Y} - { \bf X}{ \boldsymbol\Gamma} - { \bf Z}{ \boldsymbol\beta}{ \boldsymbol\alpha}^{\prime}  )\boldsymbol\Omega({ \bf Y}  - { \bf X}{ \boldsymbol\Gamma} - { \bf Z}{ \boldsymbol\beta}{ \boldsymbol\alpha}^{\prime}  )^{\prime}\Bigr)
+ \lambda_1 P_{1}(\boldsymbol\beta). \label{RRR}
\end{equation}
The above minimization problem boils down to a penalized reduced rank regression (e.g. \citealp{Chen12}). For identifiability purposes, we impose the normalization conditions 
$\boldsymbol{\alpha^{\prime}}\boldsymbol{\Omega}\boldsymbol{\alpha}=\bf I_r$. We first estimate $\boldsymbol\alpha$ conditional on  $\boldsymbol\beta$, next we estimate $\boldsymbol\beta$ conditional on  $\boldsymbol\alpha$.

For fixed $\boldsymbol\beta$, the minimization problem in \eqref{RRR} reduces to 
\begin{multline}
\hat{\boldsymbol\alpha}|\boldsymbol\Gamma_,\boldsymbol\Omega, \boldsymbol\beta  = \underset{\boldsymbol\alpha}{\operatorname{argmin}} 
\text{\ \ \ } \frac{1}{T} \text{tr} \Bigl( ({ \bf Y} - { \bf X}{ \boldsymbol\Gamma} - { \bf Z}{ \boldsymbol\beta}{ \boldsymbol\alpha}^{\prime}  )\boldsymbol\Omega({ \bf Y}  - { \bf X}{ \boldsymbol\Gamma} - { \bf Z}{ \boldsymbol\beta}{ \boldsymbol\alpha}^{\prime}  )^{\prime}\Bigr)  \text{ \ \ st. } \boldsymbol{\alpha^{\prime}}\boldsymbol{\Omega}\boldsymbol{\alpha}=\bf I_r,  \label{RRR.alpha}
\end{multline}
which is a weighted Procrustes problem \citep{Lissitz76}. This weighted Procrustes problem for $\boldsymbol \alpha$ can be seen as an unweighted Procrustes problem for $\boldsymbol{\alpha^{\star}}=\boldsymbol{\Omega^{1/2}}\boldsymbol{\alpha}$. The solution is 
\begin{equation}
\hat{\boldsymbol\alpha}= \boldsymbol\Omega^{-1/2}\boldsymbol V \boldsymbol U^{\prime}, \nonumber
\end{equation}
where $\boldsymbol U$ and $ \boldsymbol V$ are obtained from the singular value decomposition of
\begin{equation}
\boldsymbol{\beta^{\prime}}{ \bf Z}^{\prime}({ \bf Y} - { \bf X}{ \boldsymbol\Gamma})\boldsymbol{\Omega^{1/2}} = \bf U \bf D \bf V^{\prime}. \nonumber
\end{equation}
Note that \cite{Chen12} only consider the case where $\boldsymbol{\Omega}=\bf I$, and use a  Procrustes problem to solve for $\boldsymbol\alpha$. A weighted Procrustes problem takes the covariance structure into account.

\bigskip

For fixed  $\boldsymbol\alpha$, the minimization problem in \eqref{RRR} reduces to 
\begin{equation}
\hat{\boldsymbol\beta}|\boldsymbol\Gamma_,\boldsymbol\Omega, \boldsymbol\alpha  = \underset{\boldsymbol\beta}{\operatorname{argmin}} 
\text{\ \ \ } \frac{1}{T} \text{tr} \Bigl( ({ \bf Y} - { \bf X}{ \boldsymbol\Gamma} - { \bf Z}{ \boldsymbol\beta}{ \boldsymbol\alpha}^{\prime}  )\boldsymbol\Omega({ \bf Y}  - { \bf X}{ \boldsymbol\Gamma} - { \bf Z}{ \boldsymbol\beta}{ \boldsymbol\alpha}^{\prime}  )^{\prime}\Bigr) + \lambda_1 P_{1}(\boldsymbol\beta).  \label{RRR.beta}
\end{equation} 
Since $\boldsymbol{\alpha^{\star\prime}}\boldsymbol{\alpha^{\star}}=\bf I_r$, there exists a matrix $\boldsymbol{\alpha}^{\star \perp}$ with orthonormal columns such that $(\boldsymbol{\alpha}^{\star}, \boldsymbol{\alpha}^{\star \perp})$ is an orthogonal matrix. Then, with $\bf \tilde{Y}={ \bf Y} - { \bf X}{ \boldsymbol\Gamma}$,
\begin{eqnarray}
\text{\ \ \ tr} \Bigl( ({ \bf \tilde{Y}} - { \bf Z}{ \boldsymbol\beta}{ \boldsymbol\alpha}^{\prime}  )\boldsymbol\Omega({ \bf \tilde{Y}}  - { \bf Z}{ \boldsymbol\beta}{ \boldsymbol\alpha}^{\prime}  )^{\prime}\Bigr) & = &  || ({ \bf \tilde{Y}} - { \bf Z}{ \boldsymbol\beta}{ \boldsymbol\alpha}^{\prime}  )\boldsymbol{\Omega^{1/2}}||^2 \nonumber \\ 
& = & || ({ \bf \tilde{Y}}\boldsymbol{\Omega^{1/2}} - { \bf Z}{ \boldsymbol\beta}{ \boldsymbol\alpha}^{\star \prime}  )||^2 \nonumber \\ 
& = & || ({ \bf \tilde{Y}}\boldsymbol{\Omega^{1/2}} - { \bf Z}{ \boldsymbol\beta}{ \boldsymbol\alpha}^{\star \prime}  ) (\boldsymbol{\alpha}^{\star}, \boldsymbol{\alpha}^{\star \perp})||^2 \nonumber \\ 
& = & || ({ \bf \tilde{Y}}\boldsymbol{\Omega^{1/2}}\boldsymbol{\alpha}^{\star} - { \bf Z}{ \boldsymbol\beta}||^2 + || ({ \bf \tilde{Y}}\boldsymbol{\Omega^{1/2}}\boldsymbol{\alpha}^{\star \perp})||^2. \nonumber  
\end{eqnarray}
Since the second term on the left-hand-side does not involve $\boldsymbol{\beta}$, the minimization problem reduces to
\begin{equation}
\hat{\boldsymbol\beta}|\boldsymbol\Gamma_,\boldsymbol\Omega, \boldsymbol\alpha  = \underset{\boldsymbol\beta}{\operatorname{argmin}} 
\text{\ \ \ } \frac{1}{T} \text{tr} \Bigl( ({ \bf \tilde{Y}}\boldsymbol{\Omega^{1/2}}\boldsymbol{\alpha}^{\star} - { \bf Z}{ \boldsymbol\beta})({ \bf \tilde{Y}}\boldsymbol{\Omega^{1/2}}\boldsymbol{\alpha}^{\star} - { \bf Z}{ \boldsymbol\beta} )^{\prime}\Bigr) + \lambda_1 P_{1}(\boldsymbol\beta).  \label{RRR.beta2}
\end{equation}
The above minimization problem is a penalized multivariate regression of $ \left( { \bf \tilde{Y}}\boldsymbol{\Omega^{1/2}}\boldsymbol{\alpha}^{\star} \right )$ on $ { \bf Z}$.  We consider both a Lasso penalty, as in equation \eqref{L1penalty}, and an Adaptive Lasso penalty, as in equation \eqref{L1ALasso}.  
\bigskip

\noindent
{\bf Solving for $\boldsymbol\Omega$ conditional on $\boldsymbol\Gamma,\boldsymbol\Pi$.}
When $\boldsymbol\Gamma$ and $\boldsymbol\Pi$  are fixed, the minimization problem in \eqref{penloglik} is equivalent to minimizing
\begin{equation}
\widehat{\boldsymbol\Omega}| \boldsymbol\Gamma, \boldsymbol\Pi = \underset{\boldsymbol\Omega}{\operatorname{argmin}} \text{\ \ \ } \frac{1}{T} \text{tr} \Bigl( ({ \bf Y} - { \bf Z}{ \boldsymbol \Pi}^{\prime} - { \bf X}{ \boldsymbol\Gamma} )\boldsymbol\Omega({ \bf Y} - { \bf Z}{ \boldsymbol \Pi}^{\prime} - { \bf X}{ \boldsymbol\Gamma} )^{\prime}\Bigr)  - \text{log}|\boldsymbol\Omega| +  \lambda_3 P_{3}(\boldsymbol\Omega). \label{glasso}
\end{equation}
The above minimization problem corresponds to penalized covariance estimation. With the penalty term as given in equation \eqref{L1penaltyOmega}, this problem can be solved using the glasso algorithm of \cite{Friedman07}.

\bigskip

\noindent
We iterate solving minimization problem \eqref{MLM}, \eqref{RRR} and \eqref{glasso} until the angle between the estimated cointegration space in two successive iterations is smaller than some tolerance value $\epsilon$ (e.g. $\epsilon=10^{-3}$).

\bigskip

\noindent
{\it Selection of tuning parameters.}
We select the tuning parameters $\lambda_1$, controlling the penalization on the cointegrating vectors, and $\lambda_2$, controlling the penalization of the short-run effects, according to a time series cross-validation approach \citep{Rforecast}, see Appendix B. Since the sparseness structure of each cointegrating vector can be different, we allow the selected sparsity parameter $\lambda_1$ to be different for each cointegrating vector. The tuning parameter $\lambda_3$, controlling the penalization on the off-diagonal elements of $\boldsymbol\Omega$, is selected according to the Bayesian Information Criterion \citep{Friedman07}.

\bigskip

\noindent
{\it Starting values.} A starting value for $\boldsymbol\alpha$,  $\boldsymbol\beta$ and $\boldsymbol\Omega$ is required. We start with $\boldsymbol\Omega=\bf I_q$. Starting values for $\boldsymbol\alpha$ and $\boldsymbol\beta$ are obtained by first applying the iterative algorithm with an $L_2$ penalty on the cointegrating vectors, initialized by taking every component of $\boldsymbol \beta$ equal to one and $\boldsymbol \Gamma_k$ (for $k=1,\ldots, p-1)$ the identity matrices.

\bigskip

\noindent
{\it Unpenalized objective function.} The unpenalized case (i.e. $\lambda_1=0$, $\lambda_2=0$  and $\lambda_3=0$) can also be solved using the iterative algorithm. We numerically verified that this iterative procedure and Johansen's closed-form solution yield almost identical results, justifying the use of our iterative procedure to solve objective function (1).

\section{Determination of Cointegration Rank}
At small finite samples, the asymptotic distribution of Johansen's trace statistic, used to determine the cointegration rank, might poorly approximate the true distribution, resulting in substantial size and power distortions (e.g. \citealp{Johansen02};  \citealp{Nielsen04}; \citealp{Juselius06}; \citealp{Breitung09}). 
We use an iterative procedure based on the Rank Selection Criterion (RSC) of \cite{Bunea2011} to determine the cointegration rank $r$. We start with an initial value of the cointegration rank $r_{\text{start}}=q$. 

For this initial value, we first obtain $\widehat{\boldsymbol\Gamma}$, using the algorithm discussed in Section 3. In a second step, we update our estimate of the cointegration rank. Following \cite{Bunea2011}, $\hat{r}$ is given by the number of eigenvalues of the matrix ${\bf\widetilde{Y}}^{\prime}{\bf P}{\bf\widetilde{Y}}$ that exceed a certain threshold $\mu$:
\begin{equation}
\hat{r} = {\operatorname{max}} \{r : \lambda_r({\bf\widetilde{Y}}^{\prime}{\bf P}{\bf\widetilde{Y}}) \ge \mu \}, \nonumber
\end{equation}
with ${\bf\widetilde{Y}}= \bf Y - \bf X \widehat{\boldsymbol\Gamma}$ and ${\bf P}={\bf Z}({\bf Z}^{\prime}{\bf Z})^{-}{\bf Z}^{\prime}$ the projection matrix onto the column space of ${\bf Z}$. Following the recommendation of \cite{Bunea2011}, the threshold is set equal to $\mu=2S^2(q+l)$, with $l=\operatorname{rank}({\bf Z})$ and 
$$S^2=  \frac{||{\bf\widetilde{Y}} - {\bf P}{\bf\widetilde{Y}} ||^2_F}{Tq - lq},$$ 
where $||\cdot||_F$ denotes the Frobenius norm for a matrix. We repeat the above procedure using the new value of $\hat{r}$, this until the estimate of the cointegration rank does not change in two successive iterations.

The Rank Selection Criterion provides a consistent estimate of the effective rank of the coefficient matrix $\boldsymbol\Pi$ in the penalized reduced rank regression \citep{Bunea2011}. The consistency results are valid when either the length of the time series or the number of time series grows to infinity.  This procedure to determine the rank has almost no computational cost and can also be used when the number of time series is larger than the sample size.

\section{Simulation Studies}
We conduct a simulation study to evaluate the performance of the penalized ML estimator. The considered data generating process (revised from \citealp{Cavaliere12}) is the following VECM: 
\begin{equation} \label{DGP1}
\boldsymbol\Delta {\bf y}_t = \boldsymbol\alpha \boldsymbol\beta^{\prime} {\bf y}_{t-1}  + \boldsymbol\Gamma_{1} \boldsymbol\Delta {\bf y}_{t-1}   + {\bf e}_t, \ \ \ \ \ \ (t= 1, \ldots, T), \nonumber
\end{equation}
where the error terms ${\bf e}_t $ follow a $N_q(\bf{0},\bf{I}_{q})$ distribution. We set ${\bf y}_0 = \boldsymbol\Delta {\bf y}_0 = \bf{0}$. We compare the precision accuracy of the penalized ML algorithm to the maximum likelihood procedure of \cite{Johansen88}.
\bigskip

\noindent
{\bf 5.1 Simulation designs.} Different simulation designs are considered: 
(i) low-dimensional designs ($T=500, q=4$), and (ii)  high-dimensional designs with moderate sample size ($T=50, q=11$)\footnote{$q=11$ time series is the largest number for which the critical values of Johansen's trace statistic are tabulated in Johansen (Chapter 15; \citeyear{Johansen96}) or \cite{Osterwald92}. Note that \cite{Doornik98} provide a response surface approximation to the critical values tabulated by Johansen for $q$ up to at least 15.}. For each design, we consider both sparse and non-sparse simulation settings. Full details on each simulation design can be found in Appendix A. The number of simulations for each setting is $M=500$.

\textit{Low-dimensional designs.} The true cointegrating vectors are sparse in the first two simulation settings. The cointegration rank equals $r=1$, $r=2$ respectively. In the third simulation setting, the true cointegrating vector is non-sparse and $r=1$.

\textit{High-dimensional designs.} In the first two simulation settings, the true cointegrating vectors are sparse. The cointegration rank equals $r=1$, $r=4$  respectively. In the third simulation setting, the true cointegrating vector is non-sparse and $r=1$.
\bigskip

\noindent
{\bf 5.2 Performance measures.}
To evaluate the estimation accuracy, we compute for each simulation run $m$, with $m=1,\ldots,M$, the angle $\theta^{(m)}(\hat{\boldsymbol\beta}^{(m)},\boldsymbol\beta)$ between the estimated cointegration space and the true cointegration space. The average angle is then given by
\begin{equation}
\theta(\hat{\boldsymbol\beta},\boldsymbol\beta) = \dfrac{1}{M} \sum_{m=1}^{M}\theta^{(m)}(\hat{\boldsymbol\beta}^{(m)},\boldsymbol\beta).
\end{equation}

Furthermore, we evaluate the performance of the Rank Selection Criterion to the trace statistic of \cite{Johansen88}, the Bartlett-corrected trace statistic \citep{Johansen02} and the bootstrap procedure of \cite{Cavaliere12} in correctly selecting the true cointegration rank.\footnote{All tests are conducted at the 5\% significance level.} 
The Bartlett-corrected trace statistic \citep{Johansen02} and bootstrap procedure are used to improve the small sample performance of Johansen's trace statistic. 
For each method, we record the relative frequencies, over all simulation runs, of the selected ranks.
\bigskip
 
\noindent
{\bf 5.3 Results for the low-dimensional designs.}
The simulation results on the accuracy of the estimated cointegration space are reported in Table \ref{EstimAccuracy_4dim}. For different values of the adjustment coefficients, we report the average angle (averaged across simulation runs) between the estimated and the true cointegration space. We use a two-sided paired $t$-test to test equality of the average angle of the sparse estimation method and of Johansen's method. 

\begin{table}
\footnotesize
\begin{center}
\caption{\small Low-dimensional designs: $T=500, q=4$.  Average angle between the estimated and true cointegration space. The results are reported for different values of the adjustment coefficient $a$. Significant differences, at the 5\% significance level, between the sparse method and Johansen's method are in bold. \label{EstimAccuracy_4dim}}
\begin{tabular}{lllccccccccc} \hline
 Method $\backslash$ $a$ &&&&&   $-0.2$ && $-0.4$ && $-0.6$ && $-0.8$  \\ \hline
 \multicolumn{5}{l}{\it \bf  Sparse $\boldsymbol \alpha$ and $\boldsymbol \beta, r=1$ } &&&&&&& \\
\multicolumn{3}{l}{Johansen} &&& 0.060 && 0.032 && 0.020 && 0.015\\
\multicolumn{3}{l}{Sparse Lasso} &&& \textbf{0.034} && \textbf{0.018} && \textbf{0.012} && \textbf{0.009} \\ 
\multicolumn{3}{l}{Sparse Adaptive Lasso} &&&  \textbf{0.011} && \textbf{0.004} && \textbf{0.003} && \textbf{0.002}  \\
  &&&&&&&&&&& \\
\multicolumn{5}{l}{\it \bf Sparse $\boldsymbol \alpha$ and $\boldsymbol \beta, r=2$ } &&&&&&& \\
\multicolumn{3}{l}{Johansen} &&& 0.013 && 0.006 && 0.004 && 0.003\\
\multicolumn{3}{l}{Sparse Lasso} &&&  \textbf{0.007} && \textbf{0.003} && \textbf{0.002} &&\textbf{0.002}\\ 
\multicolumn{3}{l}{Sparse Adaptive Lasso} &&&\textbf{0.001} && \textbf{0.001} && \textbf{0.001} &&\textbf{0.001}\\
     &&&&&&&&&&& \\ 
\multicolumn{5}{l}{\it \bf Non-sparse $\boldsymbol \alpha$ and $\boldsymbol \beta, r=1$ } &&&&&&& \\
\multicolumn{3}{l}{Johansen} &&& 0.026 && 0.013 && 0.009 && 0.007 \\
\multicolumn{3}{l}{Sparse Lasso} &&&  \textbf{0.073} && 0.013 && 0.009 && 0.007 \\ 
\multicolumn{3}{l}{Sparse Adaptive Lasso} &&& \textbf{0.077} && 0.014 && 0.009 && 0.007\\  \hline
\end{tabular}
\end{center}
\end{table}
\normalsize

In the sparse settings, the sparse methods are the best performing. They provide significantly more precise estimates than the Johansen procedure. For almost all values of the adjustment coefficients, the estimation accuracy of the sparse methods is even twice as good as that of Johansen's method. The Sparse Adaptive Lasso provides a more precise estimate of the cointegration space than the Sparse Lasso.
In the non-sparse setting, Johansen's method is best performing for low values of the adjustment coefficients. For higher values of $a$, however, all methods show similar performance. The usage of the sparse procedures does not lead to an lower estimation precision here.  
     
Table \ref{Rank_4dim} reports the results on the determination of the cointegration rank. For reasons of brevity, we only report the results for $a=-0.4$ and $a=-0.8$. 
In the first sparse design, the Rank Selection Criterion achieves competitive performance with a rank recovery percentage around $91\%$. Note that Johansen's method is aimed at controlling size, resulting in a rank recovery percentage around 95\% when working with a 5\% significance level. 
In the second sparse design, RSC is the best performing method. It correctly selects the cointegration rank in almost all simulation runs. In the non-sparse design, Johansen's procedure performs best. The rank recovery percentage of RSC remains close to that of Johansen's trace statistic.
\bigskip

\begin{table}
\footnotesize
\begin{center}
\caption{\small Low-dimensional designs: $T=500, q=4$. Frequency of the estimated cointegration rank $\hat{r} = 0,\ldots, q$ using Johansen's trace statistic, the Bartlett-corrected trace statistic, the bootstrap of \cite{Cavaliere12} and the Rank Selection Criterion (RSC). \label{Rank_4dim}}
\begin{tabular}{lllcccccccccccc}
\hline
True rank & Method $\backslash$ $\hat{r}$ && 0 & 1 & 2 & 3 & 4 &&& 0 & 1 & 2 & 3 & 4 \\ \hline
\multicolumn{4}{l}{{\it \bf  Sparse $\boldsymbol \alpha$ and $\boldsymbol \beta$ } } && $\boldsymbol{a=-0.4}$ &&&&&&& $\boldsymbol{a=-0.8}$ &&\\
$r=1$ & Johansen&& 0.0 & 95.8  & 3.8 &  0.4  &   0.0   &&&  0.0 & 95.8  & 4.0 &  0.2 &    0.0\\
& Bartlett		 && 0.0 & 95.0 & 4.2 &  0.8  &  0.0   &&&  0.0 & 95.4 & 4.0 & 0.6 &   0.0\\
& Bootstrap 	 &&  0.0 & 96.2  & 3.4 &  0.4   & 0.0  &&&  0.0 & 96.0  & 3.8 &  0.2 &    0.0\\
& RSC            && 0.0 & 91.0  & 9.0 &  0.0    & 0.0   &&&  0.0 & 91.6  & 8.4 &  0.0 &   0.0   \\
& &&&&&&&&&&&&&\\
\multicolumn{4}{l}{{\it \bf Sparse $\boldsymbol \alpha$ and $\boldsymbol \beta$ } }&& $\boldsymbol{a=-0.4}$ &&&&&&& $\boldsymbol{a=-0.8}$ &&\\
$r=2$ & Johansen&& 0.0  & 0.0 & 96.4  & 3.6  & 0.0  &&&  0.0  & 0.0 & 96.0  & 3.8  & 0.2\\
& Bartlett		 && 0.0 &    0.0 & 93.0 & 6.8 & 0.2   &&&  0.0 &    0.0 & 95.2 & 4.8 & 0.0\\
& Bootstrap 	 && 0.0  & 0.0 & 96.0  & 3.6  & 0.4   &&&  0.0  & 0.0 & 95.4  & 4.2  & 0.4\\
& RSC            && 0.0  & 0.0 & 99.6  & 0.4  & 0.0   &&&  0.0  & 0.0 & 99.8  & 0.2  & 0.0\\
& &&&&&&&&&&&&&\\
\multicolumn{4}{l}{{\it \bf Non-sparse $\boldsymbol \alpha$ and $\boldsymbol \beta$ } }&& $\boldsymbol{a=-0.4}$ &&&&&&& $\boldsymbol{a=-0.8}$ &&\\
$r=1$ & Johansen&&  0.0 & 94.6  & 5.0  & 0.4  &  0.0   &&&  0.0 & 95.6  & 3.6  & 0.8 &   0.0\\
& Bartlett		 && 0.0 & 95.6  &  4.0 &  0.4  &  0.0   &&&  0.0 & 95.6 & 3.8 & 0.6 &   0.0\\
& Bootstrap 	 && 0.0 & 95.6  & 4.2  & 0.2  &  0.0  &&&  0.0 & 96.0  & 3.4  & 0.6 &   0.0\\
& RSC            && 0.0 & 90.4  & 9.6  & 0.0  &  0.0   &&&  0.0 & 91.4  & 8.6  & 0.0 &   0.0 \\
\hline
\end{tabular}
\end{center}
\end{table}
\normalsize
\noindent
{\bf 5.4 Results for the high-dimensional designs.}
In these designs, we expect that the advantage of using the sparse procedures becomes much larger. The sample size is small compared to the number of time series, such that the estimation imprecision when using Johansen's approach will become large. The simulation results on the estimation accuracy of the estimated cointegration space are reported in Table \ref{EstimAccuracy_11dim}. In all settings, the sparse procedures indeed significantly outperform Johansen's procedure. Also for the non-sparse design the sparse estimation procedures perform best. The differences are outspoken. Since the Lasso and Adaptive Lasso perform regularization, their good performance is retained in non-sparse high-dimensional settings. Furthermore, the Sparse Lasso and the Sparse Adaptive Lasso show similar performance.

\begin{table}
\footnotesize
  \begin{center}
  \caption{\small High-dimensional designs: $T=50, q=11$.  Average angle between the estimated and true cointegration space. Results are reported for different values of the adjustment coefficient $a$. Significant differences, at the 5\% significance level, between the sparse method and Johansen's method are in bold. \label{EstimAccuracy_11dim}}
  \begin{tabular}{lllccccccccc} \hline
  Method $\backslash$ $a$ &&&&&  $-0.2$ && $-0.4$ && $-0.6$ && $-0.8$ \\ \hline
  \multicolumn{5}{l}{{\it \bf Sparse $\boldsymbol \alpha$ and $\boldsymbol \beta, r=1$ } } &&&&&&&\\
  \multicolumn{3}{l}{Johansen} &&& 1.203 && 1.025 &&  0.825 && 0.672 \\
  \multicolumn{3}{l}{Sparse Lasso} &&& \textbf{0.791} && \textbf{0.396} && \textbf{0.228} && \textbf{0.099} \\
  \multicolumn{3}{l}{Sparse Adaptive Lasso} &&& \textbf{0.816} && \textbf{0.392} && \textbf{0.209} && \textbf{0.090} \\
        &&&&&&&&&&& \\
  \multicolumn{5}{l}{{\it \bf  Sparse $\boldsymbol \alpha$ and $\boldsymbol \beta, r=4$ } } &&&&&&&\\
  \multicolumn{3}{l}{Johansen} &&& 0.184 && 0.101 && 0.064 && 0.047 \\
  \multicolumn{3}{l}{Sparse Lasso} &&& \textbf{0.154} && \textbf{0.076} && \textbf{0.047} && \textbf{0.034} \\
  \multicolumn{3}{l}{Sparse Adaptive Lasso} &&&  \textbf{0.152} && \textbf{0.070} && \textbf{0.042} && \textbf{0.033} \\
          &&&&&&&&&&& \\
  \multicolumn{5}{l}{{\it \bf  Non-sparse $\boldsymbol \alpha$ and $\boldsymbol \beta, r=1$ } } &&&&&&&\\
  \multicolumn{3}{l}{Johansen} &&& 1.203 && 1.005 && 0.810 && 0.656 \\
  \multicolumn{3}{l}{Sparse Lasso} &&& \textbf{0.730} && \textbf{0.384} && \textbf{0.250} && \textbf{0.161} \\
  \multicolumn{3}{l}{Sparse Adaptive Lasso} &&&  \textbf{0.758} && \textbf{0.403} && \textbf{0.266} && \textbf{0.174} \\ \hline
  \end{tabular}
  \end{center}
  \end{table}
  \normalsize

\begin{table}
\scriptsize
\begin{center}
\caption{\small High-dimensional designs: $T=50, q=11$. Frequency of the estimated cointegration rank $\hat{r} = 0,\ldots, q$. \label{Rank_11dim}}
\begin{tabular}{lllcccccccccccc}
\hline
True rank & Method $\backslash$ $\hat{r}$ && 0 & 1 & 2 & 3 & 4 & 5 & 6 & 7 & 8 & 9 & 10 & 11 \\ \hline
\multicolumn{5}{l}{{\it \bf  Sparse $\boldsymbol \alpha$ and $\boldsymbol \beta$ } } &&&&& ${\boldsymbol a=-0.4}$ &&&&&\\
$r=1$ & Johansen&&  0.0 &  0.0 &  0.0 &  0.0 &   0.0 & 0.2 & 3.8 & 6.6 & 22.2 & 25.4 & 25.4 & 16.4 \\
& Bartlett		 &&  0.0 &    8.0 & 13.6 & 18.6 &  16.6 & 10.6 & 10.2 & 5.8 & 6.6 & 5.6&   4.0 &  0.4 \\
& Bootstrap 	 &&  89.6 &  9.2  & 1.2 &  0.0 &   0.0 & 0.0 & 0.0 & 0.0 &  0.0 &  0.0 &  0.0 &  0.0 \\
& RSC            && 3.8 &  95.2 &  1.0 &  0.0 &   0.0 & 0.0 & 0.0 & 0.0 &  0.0 &  0.0 &  0.0  &  0.0\\
&&&&&&&&&&&&&& \\
\multicolumn{5}{l}{  } &&&&& ${\boldsymbol a=-0.8}$ &&&&&\\
$r=1$ & Johansen&&  0.0  & 0.0 &  0.0 &  0.0 &    0.0 &  0.2  & 3.6 &  9.2&  22.6 &   24.0 &  23.4  &  17.0 \\
& Bartlett		 &&  0.0 &  2.6  & 13.4 & 23.6 & 14.6 & 11.8 & 13.8  & 7.2 &  5.2 &  4.8  &  3.0 &   0.0\\
& Bootstrap 	 &&  83.2 & 15.0 &  1.6 &  0.2  &   0.0 & 0.0 & 0.0 & 0.0 & 0.0 &   0.0  & 0.0  &   0.0 \\
& RSC            &&  0.0 & 94.6 &  5.4 &  0.0  &   0.0 & 0.0 & 0.0 & 0.0 & 0.0 &   0.0  & 0.0  &   0.0 \\
&&&&&&&&&&&&&& \\
\multicolumn{5}{l}{{\it \bf Sparse $\boldsymbol \alpha$ and $\boldsymbol \beta$ } } &&&&& ${\boldsymbol a=-0.4}$ &&&&&\\
$r=4$ & Johansen&&  0.0  & 0.0  & 0.0 &  0.0 & 0.4 &  2.8 & 24.8  & 22.6 &  27.2 &  12.6  & 5.8 &  3.8 \\
& Bartlett		 &&  0.0 &  1.4 & 7.2  & 13.4&  18.8 & 11.6 & 16.6  & 8.6 &  9.4 &  8.0 &   4.6 &   0.4\\
& Bootstrap 	 &&  75.2 & 21.6 &  3.0 &  0.2 &  0.0 &  0.0 &  0.0  & 0.0 &  0.0&   0.0  &  0.0  &  0.0 \\
& RSC            && 1.8 & 15.0 & 30.6 & 37.2 & 15.4  & 0.0 &  0.0 &  0.0 &  0.0 &  0.0   & 0.0 &  0.0 \\
&&&&&&&&&&&&&& \\
\multicolumn{5}{l}{  } &&&&& ${\boldsymbol a=-0.8}$ &&&&&\\
$r=4$ & Johansen&&  0.0 & 0.0 & 0.0 & 0.0 & 0.4 & 1.8 & 22.4 & 28.6 & 28.0 & 11.2 &  5.2 &  2.4 \\
& Bartlett		 &&  0.0 &    0.0 &  1.8 &  9.2 & 19.8 & 16.4 & 18.8 & 10.8 & 10.4 &    9.0 &  3.6  & 0.2\\
& Bootstrap 	 &&  21.6 & 44.4 & 25.6 & 6.8 & 1.4 & 0.0 & 0.0 & 0.2 & 0.0 & 0.0  & 0.0 &  0.0 \\
& RSC            &&   0.0 & 0.0 & 0.0 & 3.2& 96.8 & 0.0&  0.0 & 0.0 & 0.0 & 0.0 &  0.0  & 0.0 \\
&&&&&&&&&&&&&& \\
\multicolumn{5}{l}{{\it \bf Non-sparse $\boldsymbol \alpha$ and $\boldsymbol \beta$ } } &&&&& ${\boldsymbol a=-0.4}$ &&&&&\\
$r=1$ & Johansen&&  0.0 &  0.0 &   0.0 &  0.0 &  0.0 &   0.2 &  2.6 &  8.4 & 21.6 & 26.8 &   25.0 & 15.4 \\
& Bartlett		 &&  0.0 &  5.4 &  18.0 & 17.6  & 17.6 & 12.6 &  9.8 &  5.6 &  5.2 &  6.4  &  1.8 &   0.0 \\
& Bootstrap 	 &&  88.6 &  9.6 &  1.6 &  0.2 &  0.0 &  0.0 &  0.0 &  0.0  & 0.0 &  0.0  &    0.0 &  0.0 \\ 
& RSC            &&  6.4 & 92.6 &  1.0 &  0.0 &  0.0 &  0.0  & 0.0 &  0.0 &  0.0 &  0.0 &     0.0  & 0.0\\
&&&&&&&&&&&&&& \\
\multicolumn{5}{l}{  } &&&&& ${\boldsymbol a=-0.8}$ &&&&&\\
$r=1$ & Johansen&&  0.0 &  0.0 &  0.0 & 0.0  & 0.0 &  0.0 &  2.4 &  8.6  &  21.0 & 25.6 &    24.0 &  18.4\\
& Bartlett		 &&  0.0 &    2.0 & 16.8 & 19.8 & 18.4 & 12.4 & 10.4 &  4.8 &  6.4 &  4.6  &  4.4  &  0.0\\

& Bootstrap 	 && 80.4 & 17.4 &  2.0 &  0.2 &  0.0 &  0.0 &  0.0 &  0.0  &   0.0 &  0.0 &     0.0  &  0.0 \\
& RSC            && 0.0& 96.0 &  4.0  & 0.0 &  0.0 &  0.0 & 0.0 &  0.0&    0.0 &  0.0  &    0.0&   0.0 \\

\hline
\end{tabular}
\end{center}
\end{table}
\normalsize

Table \ref{Rank_11dim} reports the results on the determination of the cointegration rank. In all simulation designs, the Rank Selection Criterion does much better than its alternatives. In the first design with $a=-0.8$, for instance, RSC estimates the cointegration rank correctly in $94.6\%$ of the simulation runs, the Bartlett-corrected trace statistic in $2.6\%$, the bootstrap in $15.0\%$ and Johansen's trace statistic in 0\% of the simulation runs. Due to the severe size distortions in this small sample size design, the rank recovery percentage of Johansen's trace statistic does not improve when working with a significance level of, for instance, 1\%. 

When the true cointegration rank becomes higher ($r=4$ in the second design), the performance of the Rank Selection Criterion becomes sensitive to the strength of the cointegration signal: its rank recovery percentage increases from 15.4\% for $a=-0.4$ to 96.8\% for $a=-0.8$. However, even then, RSC is still the best performing method.

\section{Application}
We consider two empirical applications. In the first application on interest rates, economic theory implies sparsity in the cointegrating vectors. Therefore, it is appealing to use the sparse cointegration technique even though standard results from Johansen's system cointegration test are also available. Secondly, we perform a forecasting exercise in a large VECM of industrial production time series. 
\bigskip

\noindent
{\bf 6.1 The term structure of interest rates.}
We use the sparse cointegration approach to investigate whether the expectations hypothesis of the term structure of interest rates (EHT) holds in practice. The EHT implies that the long-term interest rate can be expressed as an average of current and market-expected future short-term interest rates plus a constant risk premium:
\begin{equation}
r_t^{\tau} = \frac{1}{\tau} \sum_{i=0}^{\tau-1} \mathbb{E}_t r_{t+i}^{1} + C, \label{EHT}
\end{equation}
where $r_t^{\tau}$ and $r_t^{1}$ are the $\tau$-period and one-period interest rates,  $C$ is a constant term premium and $\mathbb{E}_t$ is the expectations operator conditional on public information at time $t$ (e.g. \citealp{Lanne00}). We consider $q$ interest rates $r_t^1, r_t^{\tau_2},\ldots,r_t^{\tau_q}$ with increasing time to maturity $1,\tau_2,\ldots,\tau_q$. Then equation \eqref{EHT} holds for all pairs of interest rates \{$r_t^1,r_t^{\tau_2}$\}, \{$r_t^1,r_t^{\tau_3}$\}, \ldots, \{$r_t^1,r_t^{\tau_q}$\} and we can write 
\begin{equation}
r_t^{\tau} - r_t^{1} = \frac{1}{\tau} \sum_{i=1}^{\tau-1}\sum_{j=1}^{i} \mathbb{E}_t \Delta r_{t+j}^{1} + C, \label{EHTstationary}
\end{equation}
with $\Delta r_{t+j}^{1}= r_{t+j}^{1} - r_{t+j-1}^{1}$. Since the interest rates are assumed to be $I(1)$, the first differences are stationary and, hence, the right-hand-side of equation \eqref{EHTstationary} is stationary. This implies that the left-hand-side of equation \eqref{EHTstationary} must be stationary as well. There are two cointegration implications linked to the EHT. Firstly, there should be $q-1$ cointegrating vectors in a system with $q$ interest rates of different maturity; or equivalently, one common trend (with the number of common trends $=q-r$). Secondly, the $q-1$ yield spreads between the one-period interest rate and each $n$-period interest rate span the cointegration space:
\begin{equation} 
\centering
\begin{bmatrix}
1 	& 1	& \ldots & 1  \\
-1	& 0	& \ldots & 0  \\
0	& -1& \ldots & 0  \\
\vdots	& \vdots & \vdots & \vdots  \\
0	& 0 & \ldots & -1  \\
\end{bmatrix}. \label{sparsebeta}
\end{equation}
For each cointegrating vector, the sum of the cointegration coefficients should be equal to zero (``zero-sum restriction"). Rejection of one of both implications would be considered as evidence against the EHT. 

\begin{figure}
\begin{center}
\includegraphics[height=16cm]{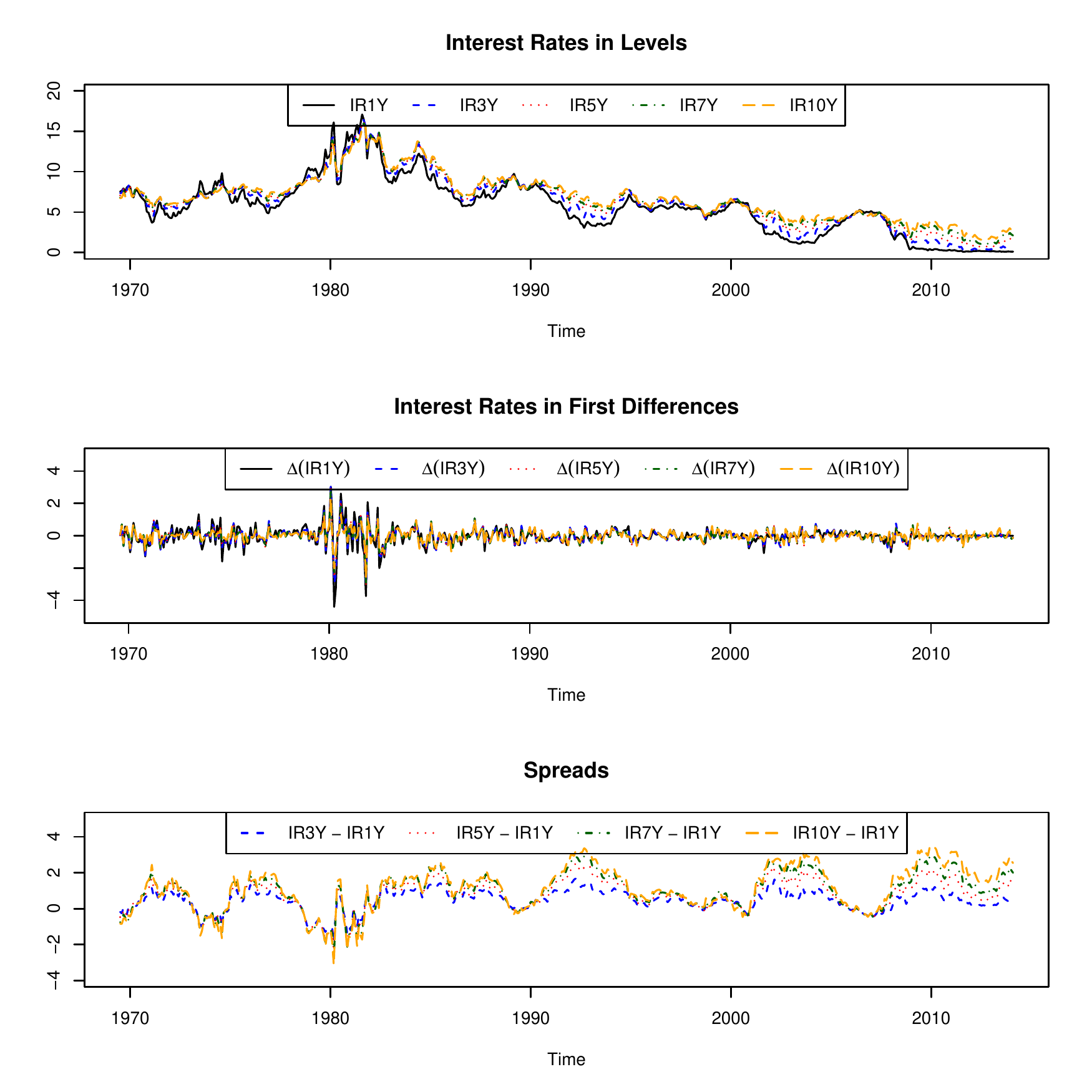}
\caption{Time plot of the interest rates (US treasury bills, constant maturity: 1-year; 3-year; 5-year; 7-year and 10-year, in \% per annum) in levels, in first differences and the spreads. Period January 1962 to February 2014. \label{IRplot}}
\end{center}
\end{figure}

We collect monthly data on five US treasury bills with different time to maturity (1, 3, 5, 7 and 10 years), ranging from July 1969 until February 2014 (source: Federal Reserve, United States). 
Time plots on the interest rates in levels, in first differences and the spreads are presented in Figure \ref{IRplot}. 
A stationarity test of all individual interest rates using the Augmented Dickey-Fuller test confirms that the time series are integrated of order 1. 

\bigskip

{ \it Cointegration Rank}. We estimate the cointegration rank using Johansen's trace statistic and the Rank Selection Criterion discussed in Section 4. Table \ref{restulsEHT} reports the results on the estimated cointegration rank. For the system with two interest rates (2-IR system), both procedures estimate the cointegration rank to be one, the number implied by the expectations hypothesis. For the other interest rate systems, the estimated cointegration rank is lower than implied by the expectations hypothesis. In the 3-IR system, for instance, both procedures underestimate the cointegration rank implied by the theory (i.e. $r=2$) by one (i.e. $\hat{r}=1$). Empirical evidence for more than one common trend is also found by other researchers. \cite{Carstensen03} and \cite{Zhang93}, for instance, find up to three common trends when interest rates of longer maturity our included. \cite{Giese08} find strong evidence for two common trends.
\bigskip

\begin{table}
\caption{Estimated cointegration rank using Johansen's trace statistic and the Rank Selection Criterion for the four interest rate systems. The last column reports the cointegration rank implied by the expectations hypothesis.  \label{restulsEHT}}
\footnotesize
\begin{center}
\begin{threeparttable}
\begin{tabular}{lcccccc} \hline
&&& \multicolumn{2}{c}{Estimated Cointegration Rank\tnote{1}}  && Expectations  \\
Interest Rate System &&& Johansen & Rank Selection Criterion &&  Hypothesis\\ \hline
2-IR system: 1Y, 2Y &&& $\hat{r}=1$ & $\hat{r}=1$ && $r=1$\\
3-IR system: 1Y, 2Y, 5Y &&& $\hat{r}=1$ & $\hat{r}=1$  && $r=2$\\
4-IR system: 1Y, 2Y, 5Y, 7Y &&& $\hat{r}=2$ & $\hat{r}=2$ &&  $r=3$\\
5-IR system: 1Y, 2Y, 5Y, 7Y, 10Y &&& $\hat{r}=3$ & $\hat{r}=2$  && $r=4$\\ \hline   
\end{tabular}
\begin{tablenotes}
\item[1] Using the Bartlett-corrected trace statistic or the Bootstrap of \cite{Cavaliere12}, we obtain the same results as for Johansen's trace statistic.
\end{tablenotes}
\end{threeparttable}
\end{center}
\end{table}

{ \it Zero-sum restriction}. We test the null-hypothesis 
\begin{equation}
H_0: \bf{ \Theta}= \begin{bmatrix} \theta_1 & \theta_2 & \ldots & \theta_{q-1} \end{bmatrix}^{\prime} = \bf{0}_{(q-1) \times 1}, \label{zerosum}
\end{equation}
with ${\theta}_j= \sum_{i=1}^{q}\beta_{ij}, \  (j=1,\ldots,q-1)$ the sum of the coefficients of the $j$th cointegrating vector. Note that the zero-sum restriction implies the cointegration space to be perpendicular to the unit vector. Therefore, every basisvector of the cointegration space needs to be perpendicular to the unit vector. 

We set the cointegration rank  $r=q-1$, the number implied by the EHT, and estimate the cointegration space using Johansen's ML procedure or the sparse penalized ML procedure  resulting in an estimate $\bf{ \widehat{\Theta}}$. To test the zero-sum restriction, 
we bootstrap the Wald test statistic $Q=\widehat{\bf{ \Theta}}^{\prime} \text{Cov}^{-1}(\widehat{\bf{ \Theta}})\widehat{\bf{ \Theta}}^{}$. Details on the bootstrap procedure can be found in Appendix C. 

Results are in Table \ref{IRspace}. Johansen's procedure reports mixed evidence. The zero-sum restriction is rejected for the 3-, 4-, and 5-IR system ($p$-values $<0.05$), but not for the 2-IR system ($p$-value $>0.05$). Estimating the cointegrating vectors with a sparse estimator, the zero-sum restriction is not rejected (for all interest rate systems $p$-values $>0.05$ ), confirming the EHT. Finally, note that many coefficients of the cointegrating vectors are estimated as exactly zero using the sparse penalized ML. This improves interpretability of the estimation results.

\begin{table}
\scriptsize
\begin{center}
\caption{Testing the zero-sum restriction for each interest rate system using Johansen's ML and Sparse penalized ML  with Lasso penalty. $P$-values are reported below every sum. \label{IRspace} }
\begin{tabular}{lccccccccccccccccccc} \hline
  &&& \multicolumn{7}{c}{Johansen ML} &&&& \multicolumn{7}{c}{Sparse Lasso} \\ 
    &&& \multicolumn{7}{c}{Cointegrating vectors} &&&& \multicolumn{7}{c}{Cointegrating vectors} \\ 
Variables &&& $\hat{\boldsymbol\beta}_1$ && $\hat{\boldsymbol\beta}_2$ && $\hat{\boldsymbol\beta}_3$ && $\hat{\boldsymbol\beta}_4$ &&&& $\hat{\boldsymbol\beta}_1$ && $\hat{\boldsymbol\beta}_2$ && $\hat{\boldsymbol\beta}_3$ && $\hat{\boldsymbol\beta}_4$\\ \hline 
&&&&&&&&&&&&&&&&&&& \\
\textbf{2-IR system} &&&&&&&&& && &&&&&&&&\\
1Y &&& 1.00 &&  && && &&&& 1.00 && && &&\\
2Y &&& -1.01 &&  && && &&&& -0.95 && && &&\\ \cline{4-5} \cline{14-15}
\textit{sum} &&& \textit{-0.01} &&  && && &&&& \textit{0.05} && && &&\\ 
&&& \multicolumn{7}{l}{$p=0.91$ } &&&& \multicolumn{7}{l}{$p=0.71$} \\ 

&&&&&&&&&&&&&&&&&&& \\
\textbf{3-IR system} &&&&&&&&& && &&&&&&&&\\
1Y &&& 1.00 && 1.00 && && &&&& 1.00 && 1.00  && &&\\ 
2Y &&& -2.41 && -8.76 && && &&&& -1.52 &&  0 && &&\\ 
5Y &&& 1.47 && 8.19 && && &&&& 0.56 && -0.87 &&  &&\\ \cline{4-7} \cline{14-17}
\textit{sum}&&& \textit{0.06} && \textit{0.43} && && &&&& \textit{0.04} && \textit{0.13} && &&\\ 
&&& \multicolumn{7}{l}{$p<0.01$} &&&& \multicolumn{7}{l}{$p=0.44$ } \\
&&&&&&&&&&&&&&&&&&& \\
\textbf{4-IR system} &&&&&&&&& && &&&&&&&&\\
1Y &&&  1.00 &&  1.00 && 1.00 && &&&& 1.00 && 1.00 && 1.00 &&  \\
2Y &&&  -19.78 && -2.00  && -6.72 && &&&&  -1.24 && -0.97  && -1.05  &&  \\
5Y &&& 48.29 &&  0.43 && 4.30  && &&&& 0 && 0  && 0.06 &&  \\
7Y &&& -29.78 && 0.64 && 1.79 && &&&& 0.27 && 0 && 0.04 &&  \\ \cline{4-9} \cline{14-19}
\textit{sum} &&&\textit{-0.27} && \textit{0.07} && \textit{0.37} && &&&& \textit{0.03} && \textit{0.03} && \textit{0.05} &&  \\ 
&&& \multicolumn{7}{l}{$p<0.01$} &&&& \multicolumn{7}{l}{$p=0.62$ } \\  

&&&&&&&&&&&&&&&&&&& \\
\textbf{5-IR system} &&&&&&&&& && &&&&&&&&\\
1Y &&& 1.00 && 1.00  && 1.00  &&  1.00 &&&& 1.00 && 1.00  && 1.00  &&  1.00   \\
2Y &&& -3.23 && 37.22  && 0.46 && -4.11 &&&& -1.34 &&  -1.07 && -1.02 && -0.80 \\
5Y &&& 1.31 && -113.98  && 0.24 && 3.51 &&&& 0   && -0.11  && 0 && 0  \\
7Y &&& 3.61 && 86.16 && -6.65 && -3.70 &&&& 0 && 0 && 0 && 0\\
10Y &&& -2.66 && -9.67 &&  5.02 && 3.61   &&&& 0.35 && 0.19 && 0  && -0.12  \\ \cline{4-11} \cline{14-20} 
\textit{sum} &&&\textit{0.03}  && \textit{0.73} && \textit{0.07} && \textit{0.31} &&&& \textit{0.01} && \textit{0.01} && \textit{-0.02} && \textit{0.08}  \\
&&& \multicolumn{7}{l}{$p<0.01$} &&&& \multicolumn{7}{l}{$p=0.11$ } \\ \hline
\end{tabular}
\end{center}
\end{table}
\bigskip

\noindent

{\bf 6.2 Forecasting industrial production in a large VECM.}
We consider a large VECM with $q=24$ industrial production time series related to manufacturing, ranging from January 1972 until January 2014. We use an updated\footnote{We extend the time range (until February 2014) and we add more industrial production time series.} version of the \cite{Stock02} database (source: Federal Reserve, United States). 
A short description of each time series can be found in Table \ref{IPVariables}, Appendix D. A stationarity test of all individual industrial production time series using the Augmented Dickey-Fuller test indicates that the time series are integrated of order one, making it appropriate to test for cointegration. 

We use the Rank Selection Criterion from Section 4 to determine the cointegration rank since it performs much better than its alternatives in the high-dimensional simulation  settings of  Section 5. The Rank Selection Criterion indicates that the industrial production time series are cointegrated with 1 cointegration equation. The sparse method includes the time series \verb|wood|, \verb|prim metal|, \verb|machinery|, \verb|electrical| , \verb|food| and \verb|non-naics|  in the cointegration equation as their associated coefficients are estimated as non-zero.

We compare the forecast performance of the Sparse penalized ML estimator (with Lasso penalty) to Johansen's ML. We estimate a VECM(1) model with one cointegration relation  for the 24 industrial production time series. We take the order of the VECM to be one, as indicated by both the Bayesian Information Criterion and the Akaike Information Criterion. Note that we have included an intercept in the VECM of equation \eqref{vecm} since some of the industrial production time series exhibit a drift.  Using a rolling window of 4 years (hence, $S=48$), the VECM is re-estimated at each time point $t=S,\ldots,T-1$ and 1-step-ahead forecasts ${\bf \hat{y}}_{t+1} = (y^{(1)}_{t+1},\ldots,y^{(24)}_{t+1})$ are computed. For each of the 24 time series ($i=1\ldots,q=24$), we compute the Mean Absolute Forecast Error (MAFE). 
\begin{equation}
\text{MAFE} = \frac{1}{T-S} \sum_{t=S}^{T-1} | \ \hat{y}^{(i)}_{t+1} -  y^{(i)}_{t+1}  \  |.
\end{equation}
Table \ref{MAE_IR} reports the results for the two forecast methods. 

\begin{table}
\caption{Mean absolute forecast error (MAFE) for the $q=24$ industrial production time series and the fwo forecast methods: Sparse penalized ML with Lasso penalty and Johansen's ML of the $q-$variate VECM with one cointegration relation. $P$-values of the Diebold-Mariano test are reported in the last column. \label{MAE_IR}}
\begin{center}
\footnotesize
\begin{tabular}{lllcccc}
  \hline
Time Series &&& Sparse Lasso & Johansen ML && $P-$value Diebold-Mariano test \\ 
  \hline
TOT &&& 1.58 & 1.74 &&  0.54 \\ 
NAICS &&& 1.58 & 1.74 &&  0.52 \\ 
DURABLE &&&  1.60 & 1.86 &&  0.34 \\ 
WOOD &&&   3.27 & 6.56 &&  $<0.01$\\ 
NONMETAL &&&  2.83 & 4.78 &&  $<0.01$\\ 
PRIM METAL &&& 4.21 & 9.99 &&  $<0.01$  \\ 
FABR METAL &&& 1.94 & 2.23 &&  0.41\\ 
MACHINERY &&& 3.25 & 4.61 &&  0.04  \\ 
COMPUTER &&&  1.14 & 0.99 &&  0.47\\ 
ELECRICAL &&& 2.95 & 5.03 &&  $<0.01$ \\ 
MOTOR &&& 4.40 & 11.51 &&  $<0.01$ \\ 
AEROSPACE &&& 3.31 & 5.22 && 0.01 \\ 
FURNITURE &&& 2.91 & 3.79 && 0.13 \\ 
OTHER DURABLE &&& 1.59 & 2.21 &&  0.02 \\ 
NONDURABLE &&& 1.61 & 2.22 &&  0.06\\ 
FOOD &&& 1.62 & 3.61 &&  $<0.01$ \\ 
TEXTILE &&& 3.50 &  7.41 &&  $<0.01$  \\ 
APPAREL &&& 4.88 & 11.69 &&  $<0.01$ \\ 
PAPER &&& 2.38 & 6.02 &&  $<0.01$\\ 
PRINT &&& 3.00 & 3.35 && 0.56\\ 
PETROLEUM &&& 2.30 & 5.59 &&  $<0.01$ \\
CHEMICAL &&& 1.89 & 2.71 &&  0.04 \\ 
PLASTIC &&& 2.45 & 3.26 &&  0.04\\ 
NON-NAICS &&& 2.78 & 3.70 &&  0.09  \\
&&&&&& \\ 
\textbf{Total} &&& 2.62 & 4.66  && $<0.01$ \\\hline
\end{tabular}
\end{center}
\end{table}

Averaged across the 24 time series, the sparse estimation procedure achieves the best forecast performance. Its forecast performance is almost two times better than that of Johansen's ML (i.e. MAFE of 2.62 against 4.66). 
Also for 23 out of 24 industrial production time series, the MAFE of the Sparse  Lasso is  lower than the MAFE of Johansen's ML. A Diebold-Mariano test confirms that the forecast performance of the Sparse Lasso is significantly, at the 5\% significance level, better than Johansen's ML, for 15 industrial production time series. In sum, we show that, in this high-dimensional application, sparsely estimating the cointegrating vector substantially improves the forecast performance compared to Johansen's approach.

\section{Conclusion}
In this paper, we discuss a sparse cointegration technique. Our simulation study shows that the sparse cointegration technique significantly outperforms Johansen's ML procedure, when the true cointegrating vectors are sparse or when the sample size is low compared to the number of time series. We  use the Rank Selection Criterion of \cite{Bunea2011} to determine the cointegration rank. In high-dimensional simulation settings, the Rank Selection Criterion  outperforms Johansen's trace statistic, the Bartlett-corrected trace statistic and the bootstrap procedure of \cite{Cavaliere12}.

Sparsity might be useful for several reasons.  First, when the underlying structure of the cointegrating vectors is known to be sparse, a sparse cointegration technique allows to explicitly capture this sparseness. We illustrate this with the expectations hypothesis. Secondly, in high-dimensional settings with cointegrated time series, estimating the cointegrating vectors with a sparse estimator might improve estimation accuracy and/or forecast performance as illustrated in the industrial production application. Third, in over-parametrized settings, where the number of time series is larger than the sample size, traditional cointegration tests can not even be computed. The sparse estimator can be used in these settings. 

There are several questions we did not address and which are left for future research. For instance, the models analyzed in this paper generally exclude deterministic terms (see e.g. \citealp{Nielsen00}). An exception is the industrial production application where an intercept was included in the VECM.  We also made abstraction of structural breaks. Allowing for structural breaks in the analysis is useful when analyzing macro-economic data \citep{Johansen00}.

A natural extension of this study would be to implement structural analysis. Impulse-response functions, for instance, can be estimated using the sparse estimator. Confidence bound around the impulse-response functions are then obtained using a bootstrap procedure. Finally, similar ideas as introduced in this paper can be used to test for Granger Causality. Few studies consider Granger Causality in high-dimensional systems, an example is \cite{Jarocinski11}. An interesting path for future research is to use a sparse procedure to test for Granger Causality in high-dimensions. 

\section*{Acknowledgments}
Financial support from the FWO (Research Foundation Flanders) is gratefully acknowledged (FWO, contract number 11N9913N).

\clearpage
\section*{Appendix A: Simulation designs}
\begin{table}[ht]
\scriptsize
\begin{center}
\caption{ \small Low-dimensional ($T=500, q=4$) and high-dimensional  ($T=50, q=11$) simulation designs.  \label{4dim}}
\begin{tabular}{l|cccccccc}  \hline
\textbf{ Low-dimensional designs}  &&&&  $\boldsymbol\beta$ && $\boldsymbol\alpha$ && $\boldsymbol\Gamma_1$ \\  \hline
&&&&&&&&\\ 
Sparse $\boldsymbol \alpha$ and $\boldsymbol \beta, r=1$ &&&& ${\boldsymbol \beta_1}=\begin{bmatrix}  1 \\ \textbf{0}_{3 \times 1}  \end{bmatrix}$ && $a{\boldsymbol \beta_1}$  && $\boldsymbol\Gamma_1=\gamma \boldsymbol I_q$ \\ 
&&&&&&&&\\ 
Sparse $\boldsymbol \alpha$ and $\boldsymbol \beta, r=2$  &&&& $ {\boldsymbol \beta_2}= \begin{bmatrix}  1 & 0 \\ 0 & 1 \\ \textbf{0}_{2 \times 1} & \textbf{0}_{2 \times 1} \end{bmatrix}$ && $a {\boldsymbol \beta_2}$ &&   $\boldsymbol\Gamma_1=\gamma \boldsymbol I_q$ \\
&&&&&&&&\\
Nonsparse $\boldsymbol \alpha$ and $\boldsymbol \beta, r=1$ &&&& ${\boldsymbol \beta_3}=\begin{bmatrix}  1 \\ 0.5 \\ 0.5 \\ 0.5 \end{bmatrix}$ && $a{\boldsymbol \beta_3}$ &&  $\boldsymbol\Gamma_1=\gamma \boldsymbol I_q$ \\
&&&& \multicolumn{5}{c}{with $a=-0.2, -0.4, \ldots, -0.8$, and $\gamma=0.1$}\\  &&&&&&&&\\   \hline

\textbf{High-dimensional designs}  &&&&  $\boldsymbol\beta$ && $\boldsymbol\alpha$ && $\boldsymbol\Gamma_1$ \\  \hline
&&&&&&&&\\ 
Sparse $\boldsymbol \alpha$ and $\boldsymbol \beta, r=1$  &&&& $ {\boldsymbol \beta_4} = \begin{bmatrix}  \textbf{1}_{3 \times 1}  \\ \textbf{0}_{8 \times 1} \end{bmatrix}$ && $a {\boldsymbol \beta_4}$ &&  $\boldsymbol\Gamma_1=\gamma \boldsymbol I_q$ \\ 
&&& &&&&&\\ 
Sparse $\boldsymbol \alpha$ and $\boldsymbol \beta, r=4$  &&&& $ {\boldsymbol \beta_5} =\begin{bmatrix}  \textbf{1}_{3 \times 1} & \textbf{0}_{3 \times 1} & \textbf{0}_{3 \times 1} & \textbf{0}_{3 \times 1}\\ \textbf{0}_{3 \times 1} & \textbf{1}_{3 \times 1} & \textbf{0}_{3 \times 1} & \textbf{0}_{3 \times 1}\\ \textbf{0}_{3 \times 1} & \textbf{0}_{3 \times 1} & \textbf{1}_{3 \times 1} & \textbf{0}_{3 \times 1} \\ \textbf{0}_{2 \times 1} & \textbf{0}_{2 \times 1} & \textbf{0}_{2 \times 1} & \textbf{1}_{2 \times 1}\end{bmatrix}$ &&$a {\boldsymbol \beta_5}$ &&  $\boldsymbol\Gamma_1=\gamma \boldsymbol I_q$ \\ 
&&& &&&&&\\
Nonsparse $\boldsymbol \alpha$ and $\boldsymbol \beta, r=1$  &&&& $ {\boldsymbol \beta_6} =\begin{bmatrix}  \textbf{1}_{3 \times 1}  \\ \textbf{0.1}_{8 \times 1} \end{bmatrix}$ &&$a {\boldsymbol \beta_6}$ &&  $\boldsymbol\Gamma_1=\gamma \boldsymbol I_q$ \\ 
&&&& \multicolumn{5}{c}{with $a=-0.2, -0.4, \ldots, -0.8$, and $\gamma=0.4$} \\  &&&&&&&&\\  \hline
\end{tabular}
\end{center}
\end{table}

\clearpage
\section*{Appendix B: Time-series cross-validation}
We select the tuning parameters according to a time series cross-validation approach \citep{Rforecast}. Denote the response by ${\bf z}_{t}$. For the penalized multivariate regression in equation \eqref{MLM}, ${\bf z}_{t}= \boldsymbol\Delta {\bf y}_t - \boldsymbol\Pi {\bf y}_{t-1}$. For the penalized reduced rank regression in equation \eqref{RRR}, ${\bf z}_{t}=\boldsymbol\Delta {\bf y}_t - \sum_{i=1}^{p-1}  \boldsymbol\Gamma_{i} \boldsymbol\Delta {\bf y}_{t-i}$. 

\small
\begin{center} 
\begin{enumerate}
\item For $t=S,\ldots,T-1$ (with $S$ such that 80\% of the data is included in the first calibration sample), repeat:
\begin{enumerate}
\item For a grid of tuning parameters, fit the model to the data  ${\bf z}_1, \ldots, {\bf z}_t$.
\item Compute the one-step-ahead forecast error ${\bf e}_{t+1}={\bf z}_{t+1} - {\bf \hat{z}}_{t+1}$ 
\end{enumerate}
\item Select the value of the tuning parameter that minimizes the mean squared forecast error
\begin{equation}
\text{MSFE} = \frac{1}{T-S} \frac{1}{q}\sum_{t=S}^{T-1} \sum_{i=1}^{q} \  \left( \frac{e^{(i)}_{t+1}}{\hat{\sigma}^{(i)}} \right )^2, \nonumber
\end{equation}
with ${e}^{(q)}_t$ the $q^{th}$ component of the multivariate time series at time $t$ and $\hat{\sigma}^{(q)}$ the standard deviation of  the time series ${z}^{(q)}_t$.
\end{enumerate}
\end{center}
\normalsize

\section*{Appendix C: Bootstrap procedure for testing the zero-sum restriction}
To test the zero-sum restriction, we use the following bootstrap procedure (see \citealp{Cavaliere12}).

\linespread{1.4}
\small
\begin{center} 
\begin{enumerate}
\item Take the cointegrating vector under the null hypothesis, $\boldsymbol\beta^{H_0}$, see equation \eqref{sparsebeta}. Given $\boldsymbol\beta^{H_0}$, use the Sparse penalized ML algorithm (or Johansen's approach)  to estimate $\hat{\boldsymbol\alpha}^{H_0} , \widehat{\boldsymbol\Gamma}^{H_0}_1, \ldots, \widehat{\boldsymbol\Gamma}^{H_0}_{p-1},$ 
together with the corresponding centered residuals $\hat{\boldsymbol\varepsilon}_{t}$.  
\item Construct the bootstrap sample recursively from
\begin{equation}\label{bootstrapdata}
\boldsymbol\Delta {\bf y}_t^{{H_0}*} = \hat{\boldsymbol\alpha}^{{H_0}} \boldsymbol\beta^{{H_0} \prime} {\bf y}_{t-1}^{*}  + \sum_{i=1}^{p-1}  \widehat{\boldsymbol\Gamma}^{{H_0}}_{i} \boldsymbol\Delta {\bf y}_{t-i}^{*}  + \boldsymbol\varepsilon_{t}^{*}, \nonumber
\end{equation}
with starting values $ { \bf y}_{t}^{*} = { \bf y}_{j}, j = 1-p, \ldots, 0 $ and with bootstrap errors $\boldsymbol\varepsilon_{t}^{*}$ obtained using a residual bootstrap such that $\boldsymbol\varepsilon_{t}^{*} =  \hat{\boldsymbol\varepsilon}_{\mathcal{U}_{t}} $ with $\mathcal{U}_{t}, t=1, \ldots, T$ an i.i.d. sequence of discrete uniform distributions on $\{1,\ldots, T\}$.
\item Apply the Sparse penalized ML algorithm (or Johansen's approach) to the bootstrap sample ${\bf y}_{t}^{H_0*}$.
\item Construct the bootstrap estimates $\widehat{\bf{ \Theta}}^{*\prime}= \begin{bmatrix}  \hat{\bf \theta}_1^* & \hat{\bf \theta}_2^* & \ldots & \hat{\bf \theta}_{q-1}^* \end{bmatrix}^{\prime}$, with $\hat{\bf \theta}^*_j= \sum_{i=1}^{q}\hat{\beta}^*_{ij}$.
\item Compute the bootstrap statistic $Q^{*}=\widehat{\bf{ \Theta}}^{*\prime} \text{Cov}^{-1}(\widehat{\bf{ \Theta}}^*)\widehat{\bf{ \Theta}}^{*}$.
\item Check if $B^{-1}  \sum_{b=1}^{B}  \bm{1}(Q^{*}_{b} >   Q) $ -  with $Q^{*}_{b}, b=1, \ldots, B$ B independent bootstrap statistics -  exceeds a fixed significance level $\eta$. If so, the null hypothesis $H_0$ is not rejected. \\
\end{enumerate}
\end{center}
\linespread{1.6}
\normalsize

\section*{Appendix D: Industrial Production Time Series}

\linespread{1.3}
\begin{table} [h]
\caption{Industrial production time series. Source: Federal Reserve, United States} \label{IPVariables}
\scriptsize
\begin{center}
\begin{tabular}{ll} \hline
Variable & Description  \\ \hline
TOT & Total manufacturing \\
NAICS & NAICS industry manufacturing \\
DURABLE  & Durable manufacturing  \\
WOOD & Wood production  \\ 
NONMETAL & Nonmetallic mineral production  \\
PRIM METAL & Primary metal  \\
FABR METAL & Fabricated metal \\
MACHINERY & Machinery \\
COMPUTER & Computer and Electronic product \\
ELECRICAL & Electrical equipment, appliance and component \\
MOTOR & Motor vehicles and parts \\
AEROSPACE & Aerospace and other miscellaneous transportation \\
FURNITURE & Furniture and related products \\
OTHER DURABLE & Miscellaneous durable manufacturing \\ 
NONDURABLE & Nondurable manufacturing \\
FOOD & Food, beverage and tobacco \\
TEXTILE & Textile and production mills \\
APPAREL & Nondurables, apparel and leather goods \\
PAPER & Paper \\
PRINT & Printing and related support activities \\
PETROLEUM & Petroleum and coal products \\
CHEMICAL & Chemical \\
PLASTIC & Plastics and rubber products \\
NON-NAICS & non-NAICS industry manufacturing \\ \hline
\end{tabular}
\end{center}
\end{table}

\clearpage
\small
\bibliographystyle{asa}
\bibliography{SparseCointegration_ref}

\end{document}